\documentclass{cs19proc}

\usepackage{kantlipsum}
\usepackage{graphicx,xspace} 
\usepackage[switch]{lineno}

\newcommand{\Sindex}{$\mathcal{S}$-index\xspace}
\newcommand{\Ssymbol}{$\mathcal{S}$\xspace}
\newcommand{\num} {$\nu_{\rm max}$\xspace}

\newcommand{\dnu} {$\Delta\nu$\xspace}
\newcommand{\Kepler} {\textit{Kepler}\xspace}

\newcommand{\Hermes} {\textsc{Hermes}\xspace}
\newcommand{\hermes} {\textsc{Hermes}\xspace}
\newcommand{\Mercator} {\textsc{Mercator}\xspace}

\newcommand{\KIC}[1]{{KIC\,#1\xspace}}

\def\teff{$T_{\mathrm{eff}}$\xspace}

\def\dnu{$\Delta\nu$\xspace}
\def\dn1{$\delta\nu_{01}$\xspace}
\def\dn2{$\delta\nu_{02}$\xspace}
\def\sun{\hbox{$_\odot$}\xspace}

\newcommand{\Sph}{$S_\mathrm{ph}$\xspace}

\usepackage[colorlinks=true,
			linkcolor=blue,
			citecolor=blue, 
			filecolor=blue, 
			urlcolor=blue]{hyperref}

\newcommand{\Figure}[1]{Figure\,\ref{#1}\xspace}

\newcommand{\Section}[1]{Section\,\ref{#1}\xspace}

\editors{G.~A. Feiden}
\publisher{Zenodo}
\conference{The 19th Cambridge Workshop on Cool Stars, Stellar Systems, and the Sun}
\conferencedate{2016}

\title{Probing seismic solar analogues through observations with  the NASA \textit{Kepler} space telescope and \Hermes high-resolution spectrograph}

\author{P.\,G. Beck,$^{1}$
 D.~Salabert,$^{1}$
 R.\,A.~Garcia,$^{1}$
 J.~do\,Nascimento, Jr.$^{2,3}$
 T.\,S.\,S.~Duarte,$^{2}$
 S.~Mathis,$^{1}$
 C.\,Regulo$^{4,5}$,\\
 J.\,Ballot$^{6,7}$,
 R.~Egeland,$^{8,9}$
 M.~Castro,$^{2}$
 F.~P{\'e}rez-Her{\'n}andez$^{4,5}$,
 O.~Creevey,$^{10}$
 A.~Tkachenko,$^{11}$
 T.~van\,Reeth,$^{11}$
L.~Bigot$^{10}$,
 E.~Corsaro,$^{1,4,5}$
 T.~Metcalfe,$^{12}$
 S.~Mathur,$^{12}$
  P.\,L.~Palle,$^{4,5}$
 C.~Allende Prieto,$^{4,5}$
 D.~Montes,$^{13}$
 C.~Johnston,$^{11}$
 M.\,F.~Andersen,$^{4,14}$  
 H.~van\,Winckel$^{11}$
 } 
 
\affiliation{
$^{1~~}$ Laboratoire AIM, CEA/DRF - CNRS - Univ. Paris Diderot - IRFU/SAp, Centre de Saclay, 91191 Gif-sur-Yvette Cedex, France\\
$^{2~~}$ Departamento de F\'isica, Universidade Federal do Rio Grande do Norte, 59072-970 Natal, RN, Brazil\\
$^{3~~}$ Harvard-Smithsonian Center for Astrophysics, Cambridge, MA 02138, USA\\
$^{4~~}$ Instituto de Astrof\'{\i}sica de Canarias, E-38200 La Laguna, Tenerife, Spain\\
$^{5~~}$ Departamento de Astrof\'{\i}sica, Universidad de La Laguna, E-38206 La Laguna, Tenerife, Spain\\
$^{6~~}$  CNRS, Institut de Recherche en Astrophysique et Plan{\'e}tologie, 14 avenue Edouard Belin, 31400 Toulouse, France\\
$^{7~~}$  Universit{\' e} de Toulouse, UPS-OMP, IRAP 31400, Toulouse, France\\
$^{8~~}$  Department of Physics, Montana State University, Bozeman, MT 59717-3840, USA\\
$^{9~~}$ High Altitude Observatory, National Center for Atmospheric Research, P.O. Box 3000, Boulder, CO 80307-3000, USA\\
$^{10~}$ Laboratoire Lagrange, Universit\'e de Nice Sophia-Antipolis, UMR 7293, CNRS, Observatoire de la C{\^ o}te d'Azur, Nice, France\\
$^{11~}$ Instituut voor Sterrenkunde, KU Leuven, B-3001 Leuven, Belgium\\
$^{12~}$ Center for Extrasolar Planetary Systems, Space Science Institute, 4750 Walnut street Suite 205, Boulder, CO 80301, USA\\
$^{13~}$ Dpto. Astrof\'{\i}sica, Facultad de CC. F\'{\i}sicas, Universidad Complutense de Madrid, E-28040 Madrid, Spain\\
$^{14~}$ Stellar Astrophysics Centre, Aarhus University, Ny Munkegade 120, 8000 Aarhus C, Denmark \\
}

\shorttitle{Ramblings of Kant}
\shortauthors{Beck et al.}

\abs{Stars similar to the Sun, known as solar analogues, provide an excellent opportunity to study the preceding and following evolutionary phases of our host star. The unprecedented quality of photometric data collected by the \Kepler NASA mission allows us to characterise solar-like stars through asteroseismology and study diagnostics of stellar evolution, such as variation of magnetic activity, rotation and the surface lithium abundance. In this project, presented in a series of papers by \cite{Salabert2016Activity,Salabert2016Mowgli} and \cite{Beck2016b,Beck2016}, we investigate the link between stellar activity, rotation, lithium abundance and oscillations in a group of 18 solar-analogue stars through space photometry, obtained with the NASA \Kepler space telescope and from currently 50+ hours of ground-based, high-resolution spectroscopy with the  \Hermes instrument. In these proceedings, we first discuss the selection of the stars in the sample,  observations and calibrations and then summarise the main results of the project.
By investigating the chromospheric and photospheric  activity of the solar analogues in this sample, it was shown that for a large fraction of these stars the measured activity levels are compatible to levels of the 11-year solar activity cycle 23. A clear correlation between the lithium abundance and surface rotation was found for rotation periods shorter than the solar value. Comparing the lithium abundance measured in the solar analogues to evolutionary models with the Toulouse-Geneva Evolutionary Code (TGEC), we found that the solar models calibrated to the Sun also correctly describe the set of solar/stellar analogs showing that they share the same internal mixing physics. Finally, the star KIC\,3241581 and KIC\,10644353 are discussed in more detail.}

\begin{document}

\maketitle


 \section{Introduction}

Solar-like oscillations are driven by the turbulent motions in the stellar convective envelope of cool stars \citep[][and references therein]{Aerts2010}. On the main sequence, these stochastic oscillations are found in stars in the spectral type range from late F to  K, spanning a mass range of roughly $\sim$1.5 down to $\sim$0.8\,M\sun \citep[e.g.][]{Chaplin2011,Chaplin2014,Huber2011}. In recent years, asteroseismic quantities were consequently used as a new tool to study and identify solar analogues in addition to the classical criteria. In the classical definition of \cite{CayreldeStrobel1996} solar analogues are seen as stars within $\pm$10\% of the solar mass, $\pm$0.3\,dex of the solar metallicity and $\pm$150\,K of the effective temperature of the Sun. The reason for the only quite recent inclusion of seismology as a criterion to search for and characterise solar-analogue stars is that high-quality and quasi-uninterrupted time-resolved data is needed to derive seismic parameters of solar-like stars.

Before the advent of the photometric space missions, such as \textsc{CoRoT}  \citep[Convection, Rotation, and planetary Transits][]{Baglin2006}, and  \textit{Kepler} \citep{Borucki2010}  such data could be only acquired from the ground through high-resolution spectrographs, with simultaneous wavelength calibration that allows us to reach meter-per-second precision (e.g. \textsc{Harps} mounted on the 3.6\,m ESO telescope at La Silla, Chile). Up to now, for only a few targets such observations could deliver rich oscillation spectra \citep[e.g. 18\,Sco][]{Bazot2012}, while   \textsc{CoRoT} and \textit{Kepler}  are providing us with vast amounts of unprecedented photometric data, allowing us to study various forms of variability throughout the Hertzsprung-Russell Diagram \citep*[HRD; for a complete overview of modern seismology see the monographs of][]{Aerts2010,TongGarcia2014,CatelanSmith2015}. 
Through these space missions, the number of solar-like oscillating main-sequence stars vastly increased to hundreds and were characterised by means of asteroseismology in numerous studies \citep[e.g.][]{Chaplin2011,Chaplin2014,Huber2011}.  

However, only a few stars that are closely matching with our Sun were found. Prime examples of  seismic analysis of analogues are 16\,Cyg\,A\&B  \citep[e.g.][]{Metcalfe2012,doNascimento2014,Davies2015,Roxburgh2016} and \textsc{CoRoT}\,102684698 \citep{doNascimento2013}. In addition, from the 197\,096 stars observed by \Kepler \citep{Mathur2016fullKeplerSample}, \cite{Salabert2016Activity} selected 18 additional analogues, which fulfilled the criteria for solar analogues of \cite{CayreldeStrobel1996}  for detailed studies. \Figure{fig:PSD} depicts the power spectral density (PSD) of KIC\,6116048, a solar analogue studied in this project.


In his fundamental paper, \cite{Skumanich1972} has shown from observations and the combined analysis of solar-like stars in the Hyades, Pleiades, Ursa Major and the Sun that stellar rotation, chromospheric activity and the surface abundance A(Li) of lithium (Li) evolve with stellar age. 
However, such studies are typically bound to stars in open clusters, as for field stars the mass is a very challenging parameter to be determined accurately. This limitation can now be overcome through asteroseismology, based on high-precision photometric time series obtained by the \Kepler satellite \citep[e.g.][]{Kallinger2010, Chaplin2011}.
Further studies have refined the gyrochronological expressions \citep[e.g.][]{Barnes2007,Garcia2014b}, stellar activity \citep[e.g.][]{Pace2012} and the lithium abundance \citep[e.g.][]{Rebolo1988,King2000,Clarke2004,Castro2016}.  Another typical challenge is that either the angular velocity or the projected surface velocity can be measured from photometry or activity variations and the rotational broadening, respectively. This observational limitation can be solved by using the asteroseismic radius \citep[e.g.][]{Beck2016b}


In this work, we are presenting the combined analysis of these 18 solar analogues in the series of papers by \cite{Salabert2016Activity,Salabert2016Mowgli} and \cite{Beck2016b,Beck2016}.
In Section\,\ref{sec:sample}, we are describing the selection of the stars in the sample, and discuss the different methods to determine the radius, mass and age that were used. An overview over the observations and calibrations is given in Section\,\ref{sec:observation}. The results of \cite{Salabert2016Activity} on photospheric and chromospheric activity are discussed in Section\,\ref{sec:DavidActivity}.
In Section\,\ref{sec:li}, we summarise the findings of \cite{Beck2016b} on rotation, stellar age and lithium, showing the strength of combining space asteroseismology with ground-based, high-resolution spectroscopy.
Two of the most interesting stars, KIC\,3241581 \citep{Beck2016} and KIC\,10644253 \citep{Salabert2016Mowgli} are discussed in Section\,\ref{sec:singleCases}. Finally, the project is summarised in Section\,\ref{sec:conclusions}.

\section{The sample of 18 solar analogues \label{sec:sample}}

\begin{figure}[t!]
\includegraphics[width=\columnwidth]{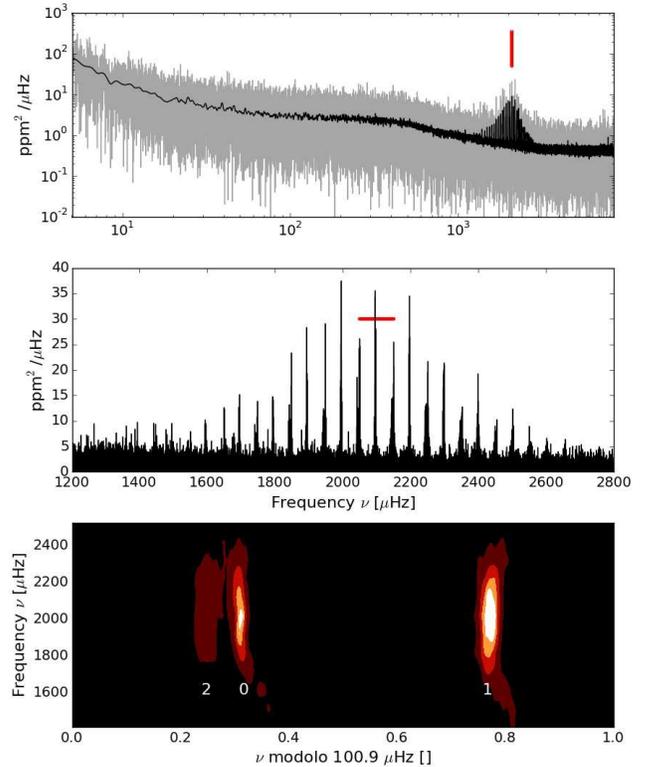}
\caption{Example of the solar analogue KIC\,61160048. The top panel shows the Power Spectral Density (PSD) in grey while a smoothed PSD is shown in black with logarithmic scaling in both axis. The centre of the excess of oscillation power is marked with a vertical red line at $\nu_{\rm max}$$\simeq$2050\,$\mu$Hz. The middle panel depicts a zoom into the frequency range of the power excess, revealing the comb-like pattern of individual oscillation modes. The large-frequency separation, \dnu, between consecutive radial modes is indicated through the horizontal red line. The bottom panel shows the \'echelle diagram that is produced by folding the PSD of the power excess with the large-frequency separation. Each vertical bright ridge corresponds to a set of oscillation modes with a given $\ell$-value.
\label{fig:PSD}}
\end{figure}

\begin{figure*}[t!]
\centering
\includegraphics[width=\textwidth,height=52mm]{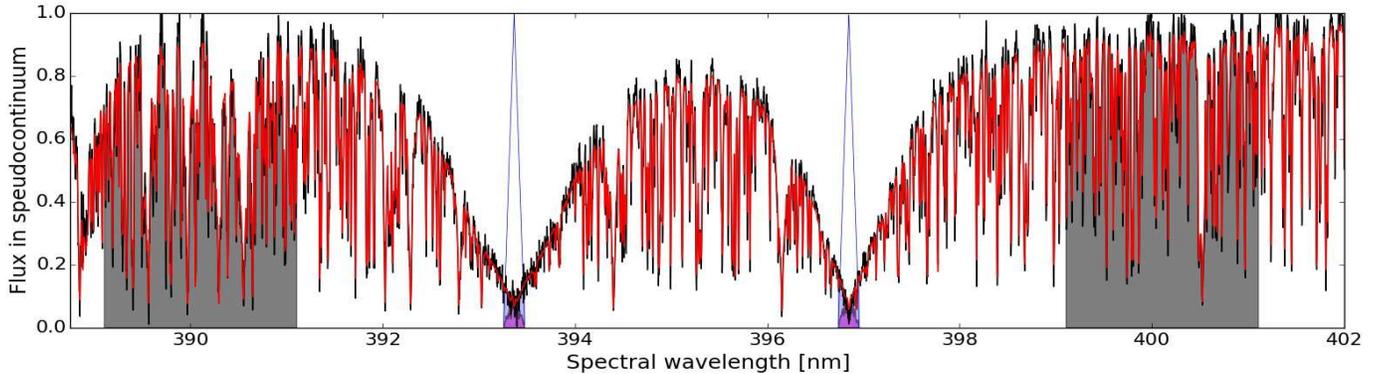}
\caption{Calculation of the \Sindex in KIC\,3241581. The blue triangle indicate the triangular filter which is used to weight the flux at the core of  the Ca\,H\&K line, left and right, respectively. It should be noted that the H line is blended with the H$_\gamma$ line.  The grey shaded regions are marking the blue and red normalisation passbands. The \Sindex is the ratio of the sum of flux from the cores of the lines, divided by the sum of the flux in the comparison passbands, times the multiplicative factor to scale the instrumental value onto the instrumental reference frame of the Mount Wilson Observatory. }
\label{fig:calculationS}
\end{figure*}

Starting from the sample of $\sim$500 main-sequence solar-like pulsating stars found in the \Kepler data by \cite{Chaplin2011}, \cite{Garcia2014b} were able to derived surface rotation periods (P$_{\rm rot}$) for $\sim$300 stars from the analysis of their light curves. Because the photometric time series, obtained by  \Kepler contain the composite signals of stellar oscillations, convection, rotational modulation and instrumental effects, \cite{Garcia2014b}, developed a combined approach of Morlet wavelet decomposition and autocorrelation techniques (a complete description of the methodology can be found in \cite{Ceillier2016}.) 

The rotation period, measured from the stellar activity is independent of the inclination of the rotation axis.
From the list of stars with determined rotation periods, \cite{Salabert2016Activity} compiled a list of 18 stars, which satisfy the criteria for solar analogues of \cite{CayreldeStrobel1996}, based on the values published in the literature. By selecting targets from a sample of \Kepler stars with known rotation periods obtained from the surface variation ensures that these stars also exhibit magnetic activity - another characteristic of our host star and not too active, because we measure oscillations \citep{Chaplin2011ApJ,Chaplin2011, Mosser2009b,Garcia2010}. 
To account for the typical uncertainty of the seismic mass estimates from individual frequencies of about 5\%, the authors expanded the mass range towards $\pm$15\% of the solar mass.


In the classical approach, stellar masses, and radii are obtained through asteroseismic scaling relations \citep[e.g.][]{Kjeldsen1995} using the centre frequency of the excess of oscillation power, \num, and the large-frequency separation, between two modes of the same spherical degree $\ell$ but consecutive orders, \dnu. Both quantities are illustrated in \Figure{fig:PSD}. Ages can be estimated from comparison of these global seismic parameters to isochrones, calculated for the corresponding stellar metallicity. For 13 stars in the sample, the masses and ages, were computed by \cite{Chaplin2014} from modelling the global seismic parameters and with constraints on temperature and metallicity from multi-colour photometry.

Stars with particularly good signal-to-noise ratios (S/N) in the oscillation spectrum 
(KIC\,3656476, KIC\,4914923, KIC\,5184732, KIC\,6116048, KIC\,7680114, KIC\,9098294, KIC\,10644253, and KIC\,3241581; \cite{Mathur2012}, \cite{Metcalfe2014}, \cite{Creevey2016}, Garcia~et~al.\ \citep[in prep.,  see also][]{Beck2016}; 
also see \Figure{fig:PSD}) were modelled with the \textit{Asteroseismic Modeling Portal} \citep[\textsc{amp},][]{Metcalfe2009} or with \textit{Modules for Experiments in Stellar Astrophysics} 
\citep[MESA,][and references therein]{MESA2013}, by using individual frequencies, frequency ratios and input parameters from high-resolution spectroscopy. This approach was critically tested and discussed by \cite{LebretonGoupil2014} who found that the realistic uncertainties of such detailed seismic modelling are $\sim$7\% and $\sim$10\% for the mass and age of the star, respectively.


\section{Space photometry \& ground-based spectroscopy \label{sec:observation}}
To study the magnetic activity of the 18 solar analogues, \cite{Salabert2016Activity} combined the analysis of the photometric and spectroscopic observations. The long-cadence photometric observations collected by the space-based \Kepler telescope were analysed over the entire duration of the mission. All the corresponding light curves were calibrated with the \textit{Kepler Asteroseismic Data Analysis and Calibration Software} \citep[KADACS,][]{Garcia2011}.

To complement space photometry, ground-based, high-resolution spectroscopic observations of these stars were obtained. In five dedicated observing runs in 2015 and 2016, we collected $\sim$53\,hrs worth of observing time for these targets with the \Hermes spectrograph \citep{Raskin2011,RaskinPhD}, mounted to the 1.2\,m \textsc{Mercator} telescope on La\,Palma, Canary Island, Spain.  The \Hermes spectra cover a wavelength range between 375 and 900\,nm with a spectral resolution of R$\simeq$85\,000. Observations for a given target were always split into several separated integrations, spread over at least 260 days. By combining the individual spectra of each star, we typically obtain a final signal-to-noise,  80$\leq$S/N$\leq$200. To be able to perform differential spectroscopic observations, a solar spectrum (S/N$\simeq$400) was obtained with the \hermes spectrum during the observing run from the solar light reflected by the Jovian moon Europa \citep{Beck2016}. We refer to \cite{Beck2016b,Beck2016} for the details of the observations.


Radial Velocities (RV) were derived for each individual observation by using the optimised line list for solar-like stars, observed with the \hermes instrument \citep{Raskin2011}. The total time span covered by our observations is typically around  300 to 350 days. The star with the longest monitoring time base is KIC\,3241581 with 700 days, which was found to be a binary by \cite{Beck2016}. Using one month of observations in 2015, \cite{Salabert2016Activity} reported that \KIC{4914923}, \KIC{7296438} and \KIC{9098294} were binary candidates. After obtaining more spectra in 2016, \cite{Beck2016b} confirmed the previous finding and further reported \KIC{10130724} and \KIC{7700968} as new binary systems. 
Because the periods are much longer than the current time span covered by our observations, no indication of the orbital period or orbital eccentricity could be inferred from the existing spectroscopy.  Several stars have RV-scatter between 200 and 300\,m/s over the time span of about a year, but without any clear trend. In principle those could be very long-periodic or plane-on binary systems. 
Our dataset is currently not suited to detect planets as this would require simultaneous wavelength calibration. In this special observing mode \Hermes is capable of a precision of 4\,m/s \citep[][]{Beck2015a} but offers a lower resolution.


\section{Photospheric \& chromospheric activity \label{sec:DavidActivity}}

\begin{figure}[t!]
\centering
\includegraphics[width=\columnwidth,height=65mm]{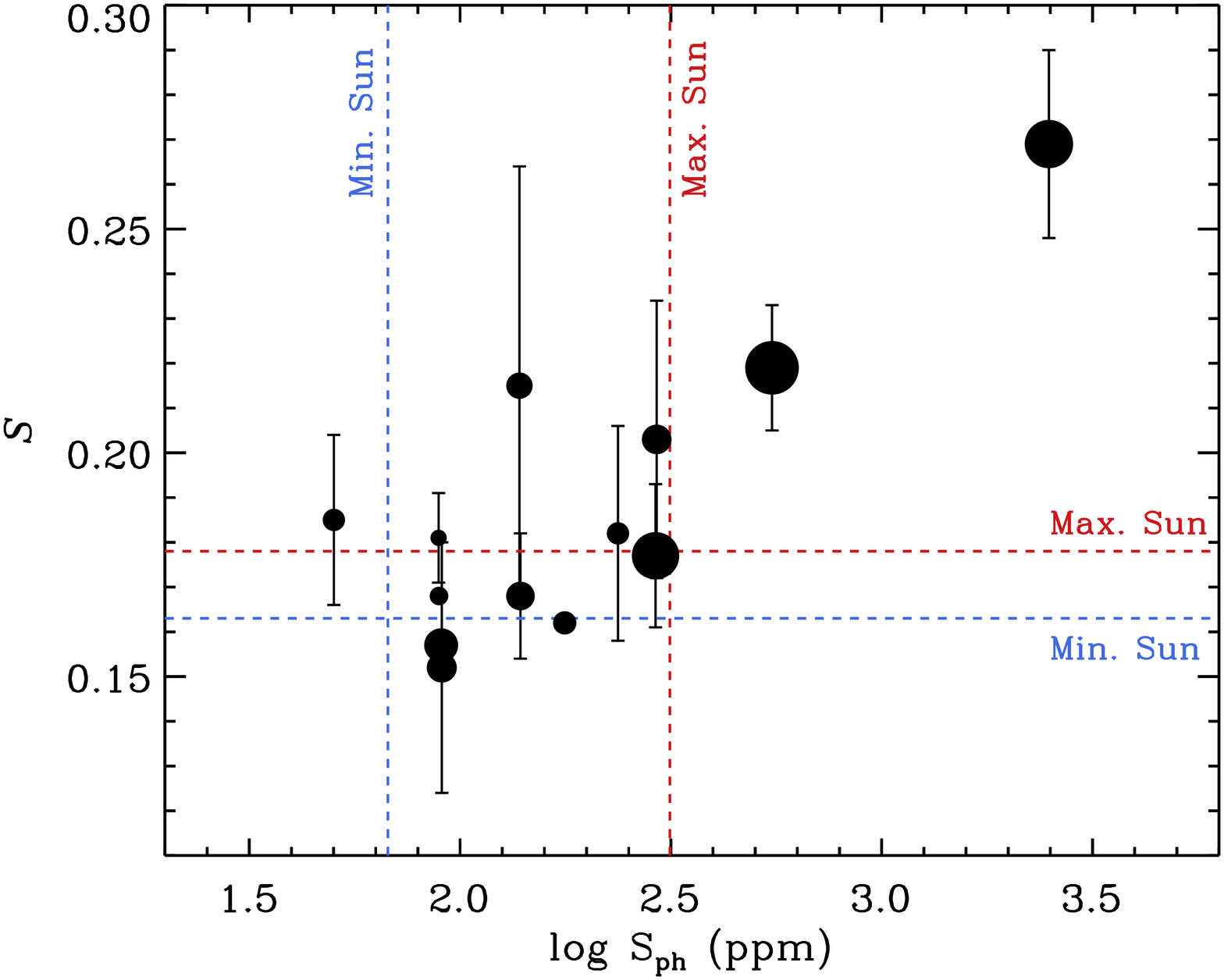}
\caption{Comparison of the photospheric \Sph-index (x axis) with the chromospheric \Sindex, calibrated into the MWO reference (y axis) for the sample of the 18 solar-analogue stars. For both \Sph and \Ssymbol, the maximum and minimum of the solar activity levels during the solar cycle are indicated by the red and blue dashed lines respectively. The size of the symbols is inversely proportional to the rotation period \citep[adopted from Figure\,3 from][]{Salabert2016Activity}.}
\label{fig:DavidSph}
\end{figure}

Studying solar analogues at various ages allows us to study how activity is changing along stellar evolution. For the Sun, this is only possible for about 10$^4$ years from the variation of the concentration of cosmogenic isotopes \citep[e.g.\ $^{10}$Be in the ice layers of the terrestrial Arctic, see][]{Hanslmeier2013}. These isotopes indicate the presence of solar activity modulation with a quasi-period of about 1000 years, in addition to the modulation from direct observations of spots and solar activity.

The study of \cite{Salabert2016Activity} quantified the properties of the magnetic activity of these 18 seismic solar analogues using the observations collected by the photometric \Kepler satellite and by the ground-based, high-resolution \Hermes spectrograph. That work aimed to test if the typical activity level of the Sun is representative for solar analogues. Such a comparison can also draw constraints on the strength of the surface magnetism and the underlying dynamo models.
The photospheric and chromospheric levels of stellar activity were measured through the \Sph and \Sindex respectively. 
The \Sph-index, which quantifies the mean variance of the light curve over subseries of length equal to five times the size of the rotation period of the star, was introduced by \cite{Mathur2014SPH} adapted from the star spot proxy proposed by \cite{Garcia2010}. The \Sindex, derived from spectroscopic observations, measures the strength of the emission in the core of the Ca\textsc{ii}\,H\&K lines in the near ultra-violet. {The strength of the emission feature is proportional to the strength of the temperature inversion between the chromosphere and the photosphere.}
The calculation of the \Sindex is illustrated and described in \Figure{fig:calculationS}. The extensive observing program from the Mount Wilson Observatory \citep[MWO,][]{Duncan1991} constitutes today the reference for the dimensionless \Ssymbol proxy. 
Indeed, the value of \Ssymbol is instrument and spectral-type dependent and thus requires a careful calibration for each instrument and object class. For the \Hermes instrument, \cite{Beck2016} determined the conversion factor to scale observations of solar-like stars into the MWO standard. 

Based on the photometric and spectroscopic data set, 
\cite{Salabert2016Activity} studied the relation between the photospheric \Sph and chromospheric \Ssymbol magnetic activity proxies and showed that they are complementary. Moreover, these two proxies provide a measure of the activity in different layers of the stellar atmosphere. They concluded that the \Sph index can be used to provide a suitable magnetic activity proxy that can be easily estimated from existing photometric observations for a large number of stars. Furthermore, they showed that 
the activity of this set of solar analogues is comparable to the solar activity within the temporal variations between minimum and maximum of the 11-year solar activity cycle  
(Figure\,\ref{fig:DavidSph}). Furthermore, \cite{Salabert2016Activity} explored further the relation between the \Sph and \Ssymbol proxies in the case of the Sun. The photospheric \Sph was calculated from the photometric observations collected by the VIRGO/SPM instrument \citep{Froehlich1995} onboard the SoHO spacecraft \citep{Domingo1995}, while the Ca K-line emission index measured at Sacramento Peak/National Solar Observatory was used to track the chromospheric activity. The comparison performed over one 11-year solar cycle indicates that the relation is not linear but present an hysteresis pattern (see their Figure\,5). 

Such hystereses have been observed in a wide range of solar observations between photospheric and chromospheric activity proxies \citep[][and references therein]{Ozguc2012}.  Nonetheless, this confirms the complementarity between \Sph and \Ssymbol derived for the solar analogues. The \Sph index thus provides an advantageous alternative to measure the stellar activity of thousands of stars without the need of investing spectroscopic observing time. However, \Sph needs to be taken as a lower limit of stellar activity as it is dependent on the inclination angle
of the rotation axis with respect to the line of sight, but it depends on the position and width of the active latitudes of the star. 
For an in-depth discussion of the astrophysical implications of this result, we refer to \cite{Salabert2016Activity}.


\begin{figure*}[t!]
\centering
\includegraphics[width=0.83\textwidth]{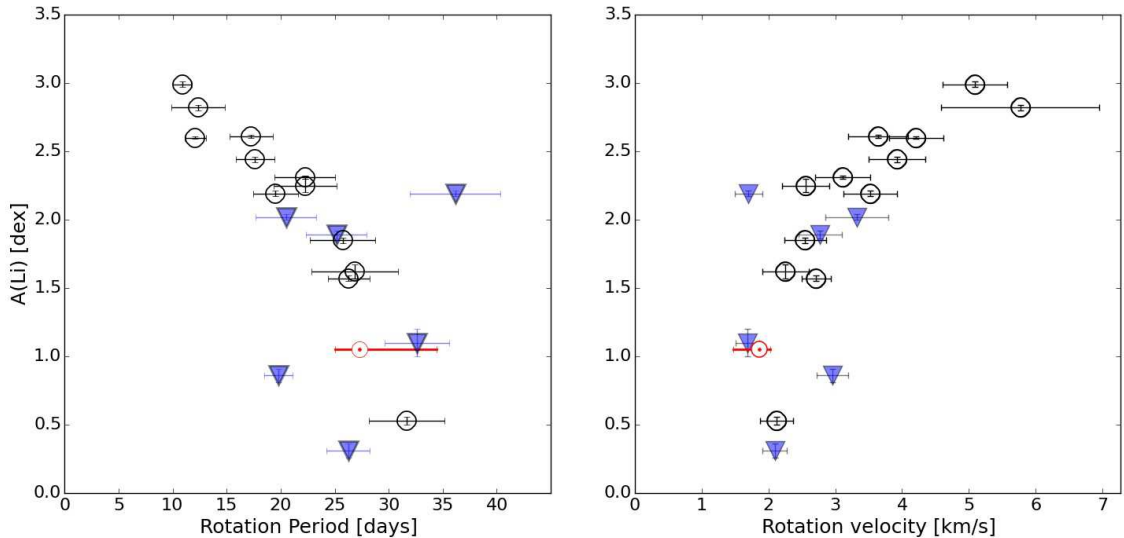}
\caption{Spectroscopic lithium abundance versus the photometric rotation period of the selected 18 solar analogues. Stars found to be located in binaries are shown as filled symbols. The range of solar rotation period as well as the period from spots are marked  through the red solar symbol \citep[patterned after Figure\,5 from][]{Beck2016}.}
\label{fig:lithiumRotation}
\end{figure*}

\section{Lithium abundance in context \label{sec:li}}

From studying isochrones in the mass-A(Li) plane for stars on the main sequence in the Hyades and other clusters, several works  identified different, mass-dependent behaviours \citep[e.g.][and references therein]{Wallerstein1965,Talon1998,Castro2016}. 
While a continuous increase of the lithium abundance is observed with rising effective temperature until \teff$\lesssim$6400\,K, it plateaus to a value close to the primordial A(Li) value that hardly changes during main-sequence phase for stars with 6400\,K$\lesssim$$T_{\rm eff}$$\lesssim$6600\,K. These temperature ranges correspond to main-sequence stars with masses of M$\lesssim$1.1M$_\odot$ and 1.1$\lesssim$M/M$_\odot$$\lesssim$1.3, respectively. Between 6600\,K$\lesssim$$T_{\rm eff}$$<$6900\,K (1.3$\lesssim$M/M$_\odot$$\lesssim$1.5) the so called Li-dip \citep{Wallerstein1965} is found, as the measured A(Li) drops towards a minimum around stars with 1.4\,M$_\odot$, before it reaches a local maximum for 6900\,K$\lesssim$\teff and slowly declines towards higher temperatures and masses.

Theoretical studies identified different physical mixing processes that are expected to work in these different temperature regimes and predict A(Li) using stellar evolution codes including rotation and macroscopic transport processes \citep[e.g.][]{Talon1998,Talon2005,Castro2016}.  For stars with M$\leq$1.1M\sun, A(Li) is mainly a function of mass.  
The angular momentum is efficiently transported through physical processes such as internal gravity waves, which are dependent on the depth of the outer convective envelope, and/or magnetic field that modify the mixing of chemicals \citep{Talon2005, Strugarek2011, Castro2016}. 
Furthermore, it was noticed by \cite{Boesgaard1987}  that the sharp drop in the surface rotation velocities around 1.5\,M\sun, known as the Kraft brake \citep{Kraft1967}, coincides with the position of the Li dip, pointing towards rotation as key parameter governing the abundance of Li \citep{Boesgaard1987}. As a conclusion, the rotationally induced mixing and internal gravity waves have a crucial role in governing the surface lithium abundance.
\citep[e.g.][respectively]{Zahn1992, Schatzman1996, MathisZahn2004,Charbonnel2005, Castro2016,Brun1999}.

 By choosing a sample of stars within the very narrow mass range of 1.0$\pm$0.10\,M\sun allows us to narrow down the sensitivity of specific effects on the mass. 
Indeed, \cite{Carlos2016} found evidence for a much stronger correlations between Li and age if instead of a wide mass range only stars close to the solar mass were analysed. However, for stars outside clusters or double-lined binaries with an accurately known inclination, a good mass estimate is a very challenging parameter to obtain. Although the stars shown in \Figure{fig:lithiumRotation} are field stars, asteroseismology provides very precise masses. 

\subsection{Li versus surface rotation and binarity}
Because for the 18 solar analogues the information on mass and radius is derived through asteroseismology and the surface rotation is measured from \Kepler photometry, they constitute a good sample to study the behaviour and dependencies of A(Li).
 \cite{Beck2016b}  studied the dependence between surface rotation, stellar age and the Li abundance. 
Comparing the surface rotation period with the lithium abundance shows a strong correlation between the two parameters for surface rotation rates, faster than the Sun (left panel of \Figure{fig:lithiumRotation}). This behaviour is in agreement with what numerous studies have found for the relation between rotation and A(Li) in clusters  \citep{Skumanich1972,Rebolo1988,King2000,Clarke2004,Bouvier2008,Bouvier2016}. 

The Li abundance can be modified by binarity. \cite{Zahn1994} showed that synchronised binaries have different lithium abundances due to their modified history of angular-momentum evolution. Furthermore, a double-lined spectroscopic binary, unnoticed due to sparse sampling, could also change the flux ratio and therefore influence the determined A(Li). Marking the binary stars, which were identified from the measured RV variation (filled symbols in \Figure{fig:lithiumRotation}), allows us to explain all stars that appear to be outliers in \Figure{fig:lithiumRotation}. 
For a complete characterisation of the systems and understanding their A(Li), we would need the orbital parameters, among which the orbital period. Because the observations do not provide a good phase coverage for the binaries in the sample yet and the stars do not show indication of binarity in their \Kepler light curve these parameters remain currently unknown.

While the rotation periods, derived from space photometry are independent of the inclination of the rotation axis, we are still facing the problem that two stars with the same rotation period do not need to have the same rotation velocity, if they have different radii. Using the value of the radius from seismology, this can be overcome by converting the rotation period into the actual rotation velocity, unaffected by projection effects along the line of sight (right panel of \Figure{fig:lithiumRotation}). It should be noted, that  by using the asteroseismic radius estimate,  we are introducing assumption on the physics of the models used to derive radius and mass such as the scaling relations or the stellar evolution code.

From the comparison of clusters and models it is known that the strongest decrease of the lithium abundance happens in the early phases of the main-sequence phase \citep[and references therein]{Castro2016}. In parallel, the surface rotation is slowing down as the star is evolving along the main-sequence \citep[e.g.][]{Barnes2007,Gallet2013,Gallet2015}.  In a recent study, \cite{vanSaders2016} showed from stellar modelling and constraints from \Kepler observations that stars at the age of the Sun or older are not braking as much as predicted by gyrochronology. This last result does not affect the found correlation between high A(Li) and fast rotation as depicted in \Figure{fig:lithiumRotation}, which shows that the gyrochronology principle is still valid for stars younger than the Sun.

\subsection{Li versus the stellar age from asteroseismology}

Because these stars are not located in clusters, their age is another challenging parameter to determine. As described in \Section{sec:sample}, asteroseismology also provides ways to  determine the stellar age. In a  further analysis, \cite{Beck2016b} focused on the four stars from the sample of the 18 that were confirmed to be single and had modelling of the individual frequency pattern available. For these stars accurate masses and ages were known from the analysis with the \textsc{amp} code, performed by \cite{Creevey2016}.

Of the four solar analogues, three formed a sequence of 'massive' solar analogues (M/M$_\odot$$\sim$1.1) with a stellar metallicity of [Fe/H]$\simeq$+0.2\,dex. Their ages range from 1 to 8\,Gyr. The fourth one has a mass of $\sim$1.05\,M\sun, but a sub-solar metallicity of [Fe/H]$\simeq$-0.2\,dex.
Furthermore, a representative grid of models predicting the temporal evolution of A(Li) due to rotation-induced mixing was calculated, using the \textit{Toulouse-Geneva stellar Evolution Code} \citep[TGEC,][]{HuiBonHoa2008, doNascimento2009} . These models were calibrated onto the solar case by \cite{Castro2016}. For a detailed description of the grid  of models we refer the reader to \cite{Beck2016}.

We compare stellar ages coming from seismology and from lithium abundances with the predicted evolution of Li from the TGEC models. It shows a good agreement within the realistic uncertainties of stellar mass and age. Although internal gravity waves were not explicitly included in the input physics of the grid of models, the calibration of the non-standard mixing onto the solar case accounted for such additional effects in a parameterised way.
Such good consensus between  theory and observations shows that in these four solar analogues, the same physical processes are driving internal mixing.


\section{Single case studies of KIC\,3241581 and KIC\,10644253 \label{sec:singleCases}}
Finally, we want to highlight two stars, that were investigated in more detail and were presented in individual papers, KIC\,324158 and KIC\,10644253. The spectroscopic time series  cover $\sim$700\,days and $\sim$400\,days, respectively. 

The star KIC\,3241581 was selected for a detailed study by \cite{Beck2016}, because it showed one of the closest match to the solar global seismic parameters among the sample of $\sim$500 \Kepler stars analysed by \cite{Chaplin2011}. Also the averaged rotation period of this star is $\sim$26 days \citep{Garcia2014b}, which implies that this star would be at around half of its main-sequence evolution, assuming general gyrochonology relations. To perform a detailed spectroscopic differential analysis of KIC\,3241581 with respect to the solar spectrum, we also obtained reference spectra, by observing the solar light, reflected by the Jovian moon Europa with the \Hermes spectrograph, during the monitoring of the target star.  The question which motivated this detailed analysis was, if a star that resembles the Sun in terms of rotation period and seismic parameters can it also be a spectroscopic solar analogue. From a spectroscopic analysis and the \Sindex (see \Figure{fig:calculationS}), 
\cite{Beck2016} found that KIC\,3241581 is a metal rich solar analogue and exhibits a typical solar activity level. This study also reported that KIC\,3241581 is a binary with an orbital period of at least 1.5 years. Although monitoring continued, \cite{Beck2016b} could not resolve the orbit from nearly two years, indicating an orbital period, at least of 3 years.

As discussed before, spots indicate magnetic fields, which is known to also be shifting the frequency of the oscillation modes at a level of a few tens of a $\mu$Hz with respect to the mean mode frequency \citep{Salabert2015}, demonstrating that seismology is sensitive to sub-surface solar activity. With the exception of the Sun \citep[e.g.][]{Salabert2015} and HD49933 \citep{Garcia2010}, no such temporal frequency variation was found in solar-like oscillators. In a detailed study of the young solar analogue KIC\,10644253, \cite{Salabert2016Mowgli} could prove the variation of the frequency of radial and non-radial modes with a timescale of about 1.5 years and generally follows the variation of the \Sph-index. Spectroscopy shows that this star has indeed a high level of chromospheric activity and a high surface abundance of lithium.
In the meantime, further detections of the seismic signature of ongoing magnetic variations were reported in additional stars \citep[]{Regulo2016, Kiefer2016}.

\section{Discussion \& Conclusions \label{sec:conclusions}}

The aim of the project on our solar-analogue stars observed with the NASA \Kepler satellite is to test, if the Sun is a typical 1\,M\sun star. Aspects of stellar structure, dynamics and evolution, such as surface rotation, stellar activity, binarity and the lithium abundance are discussed in a series of papers by \cite{Beck2016b,Beck2016}, and \cite{Salabert2016Activity,Salabert2016Mowgli}, for which the main results were summarised in these proceedings. 
For field stars this sample is unique in the sense that we combine the existing information about the rotation period, lithium abundance, seismic  masses,  and binarity status. Including high-precision asteroseismology is a tool of great help to this investigation, as it provides us with accurate estimates on several parameters (mass and age) that otherwise are only accessible for stars in well studied clusters.

The main results of the presented studies are that most of the stars exhibit solar-like activity levels. Also the new photometric activity index \Sph is a valid alternative to the \Sindex to gauge the magnetic activity level. It was also shown that for stars younger than the Sun, a clear correlation between the abundance of lithium and the rotation rate exists. By comparing the measured A(Li) from spectroscopy and derived stellar age from seismology to the prediction of the evolution of lithium by a model calibrated to the solar case and calculated for the mass and metallicity of well studied solar analogues, it was shown that these stars share the same internal mixing physics than the Sun. This suggests that, the value of the solar lithium is not peculiar.
Yet the number of stars, well characterised through photometry, spectroscopy and asteroseismology is small. New targets will be provided by space photometry through the ongoing K2 mission \citep{K2Mission}, the forthcoming missions \textsc{Tess} \citep{TESS2010} and \textsc{Plato} \citep{plato2011,Rauer2014} or from ground-based high-resolution spectroscopy with the \textsc{Song}-network \citep{Song2011}. A larger number of such well constraint targets will be needed on the way to better constrain the solar-stellar connection \citep{Brun2015}. \\

\section*{Acknowledgments}
We acknowledge the work of the team behind \Kepler and \textsc{Mercator}.  
PGB and RAG acknowledge the ANR (Agence Nationale de la Recherche, France) program IDEE (n$^\circ$~ANR-12-BS05-0008) "Interaction Des Etoiles et des Exoplanetes". PGB and RAG also received funding from the CNES grants at CEA. JDN, MC, and TD acknowledge the CNPq and the PPGF/UFRN. 
DS and RAG acknowledge the financial support from the CNES GOLF and PLATO grants.
AT is a Postdoctoral Fellow of the Fund for Scientific Research (FWO), Flanders, Belgium. 
StM acknowledges support by the ERC through ERC SPIRE grant No. 647383.
SaM would like to acknowledge support from NASA grants NNX12AE17G and NNX15AF13G and NSF grant AST-1411685.
DM acknowledges financial support from the Spanish Ministerio de Econom{\'i}a y Competitividad under grant AYA2014-54348- C3-3-R. 
The research leading to these results has received funding from the European Community's Seventh Framework Programme ([FP7/2007-2013]) under grant agreement No. 312844 (SPACEINN) and under grant agreement No. 269194 (IRSES/ASK). The observations are based on spectroscopy made with the \Mercator Telescope, operated on the island of La Palma by the Flemish Community, at the Spanish Observatorio del Roque de los Muchachos of the Instituto de Astrof{\'i}sica de Canarias. 
This research has made use of the SIMBAD database, operated at CDS, Strasbourg, France.\\

\bibliographystyle{cs19proc}
\bibliography{bibliographyLithium.bib}

\end{document}